# Classical to Quantum Diffusive Transport in Atomically Thin Semiconductors Capped with High-κ Dielectric

*Jarosław Pawłowski [1], Dickson Thian[2], Repaka Maheswar[2], Chai Jian Wei[2], Pawan Kumar[2], Sudhiranjan Tripathy[2], Hugh O.H. Churchill[3,4,5], Dharmraj Kotekar Patil[3,4,5]*

[1] Institute of Theoretical Physics, Wrocław University of Science and Technology, Wrocław, Poland

[2] Institute of Materials Research and Engineering, Agency for Science Technology and Research, (A*STAR) Singapore 138634, Republic of Singapore 138634

[3] Institute for Nano Science and Engineering, University of Arkansas, Fayetteville, AR 72701 USA

[4] Department of Physics, University of Arkansas, Fayetteville, AR 72701 USA

[5] MonArk NSF Quantum Foundry, University of Arkansas, Fayetteville, AR 72701 USA

## Abstract

The dielectric environment surrounding semiconductors plays a crucial role in determining device performance, a role that becomes especially pronounced in atomically thin semiconductors where charge carriers are confined within a few atomic layers and strongly interact with their surroundings. High-κ dielectrics, such as hafnium oxide ($HfO_2$), have been shown to enhance the performance of two-dimensional (2D) materials by suppressing scattering from charged impurities and phonons, but most studies to date have focused on room-temperature transistor operation. Their influence on quantum transport properties at low temperatures remains largely unexplored. In this work, we investigate how capping monolayer molybdenum disulfide ($MoS_2$) with $HfO_2$ modifies its electronic behavior in the quantum regime. By comparing devices with and without $HfO_2$ capping, we find that uncapped devices exhibit transport dominated by classical diffusive scattering, whereas capped devices show clear Fabry–Pérot interference patterns, providing direct evidence of phase-coherent quantum transport enabled by dielectric screening. To gain further insight, we develop a tight-binding interferometer model that captures the effect of dielectric screening on conductive modes and reproduces the experimental trends. Our findings demonstrate that dielectric engineering provides a powerful route to control transport regimes in TMD devices.

# Introduction

Two-dimensional semiconductors such as monolayer $MoS_2$ have emerged as technologically promising materials due to their unique combination of electronic and quantum properties [1], [2], [3], [4]. The presence of a sizable direct bandgap enables high on/off ratios [5], [6] in ultra-scaled transistors, while strong excitonic effects and spin–valley locking open new avenues for electronics, optoelectronics and quantum devices [6], [7], [8], [9], [10], [11], [12], [13]. These intrinsic features, together with the ability to integrate $MoS_2$ into van der Waals heterostructures without lattice matching, provide a versatile platform for developing scalable, low-power, and multifunctional device technologies [13], [14], [15]. As a result, $MoS_2$ stands out as a key candidate for next-generation electronic, photonic, and quantum devices.

The atomically thin body of two-dimensional semiconductors confines charge transport to only a few atomic layers, making their electronic properties highly sensitive to the surrounding environment. In electronic devices, the gate dielectric sits in direct contact with the channel material and hence may have impact in altering transport characteristics, offering an additional lever to control device performance. High-κ dielectrics such as hafnium oxide ($HfO_2$) have been shown to improve device operation by suppressing impurity scattering and enhancing carrier mobility, in agreement with theoretical predictions. For instance, capping monolayer $MoS_2$ with $HfO_2$ significantly reduces charged impurity scattering, leading to mobility enhancement by orders of magnitude [6], [16]. At the same time, dielectric layers, often deposited by atomic layer deposition (ALD), can introduce defects and charge transfer at the interface, resulting in unintentional doping and higher current densities [17], [18].

$HfO_2$ has therefore emerged as a crucial high-κ dielectric for advancing the performance of two-dimensional (2D) transition metal dichalcogenide (TMD) field-effect transistors (FETs), particularly single-layer molybdenum disulfide ($MoS_2$), owing to its superior gate control, low leakage current, and compatibility with existing semiconductor fabrication processes. Previous studies have shown that the dielectric environment strongly influences the electronic transport property of single layer $MoS_2$ devices. For example, devices with a vacuum/$MoS_2$/$SiO_2$ stack exhibit strong Coulomb impurity scattering due to charge traps at the $SiO_2$ interface, limiting room-temperature mobility to below ∼50 cm²/Vs [16]. Introducing a high-κ dielectric such as $HfO_2$ in an $HfO_2$/$MoS_2$/$SiO_2$ configuration has been predicted to enhance Coulomb screening, thereby improving mobility and reducing hysteresis effects [19]. While low-κ hexagonal boron nitride has historically been the dielectric of choice due to its clean interface, studies of high-κ dielectrics in two-dimensional materials particularly at deep cryogenic temperatures where quantum devices operate remain limited. Understanding how high-κ dielectrics influence electron transport, under these conditions is therefore essential for advancing both scalable nanoelectronics and next-generation quantum technologies.

Here, we report low-temperature transport studies in single-layer $MoS_2$ devices capped with $HfO_2$ and compare the results with a control device without a high-κ dielectric. In the uncapped device, transport is dominated by classical diffusive behavior in the single electron tunneling regime, whereas in the $HfO_2$ capped device, transport is governed by a quantum diffusive regime, as evidenced by quantum interference effects. To validate these observations and gain deeper insight into the underlying mechanisms, we performed simulations that show excellent agreement with the experimental measurements.

# Results

In a field-effect transistor, studying carrier mobility as a function of temperature provides insight into the underlying scattering mechanisms. At high temperatures, mobility is primarily limited by phonon scattering ($\mu_{ph}$), while at low temperatures impurity scattering dominates ($\mu_{imp}$). At an intermediate crossover temperature ($T_t$), both mechanisms contribute equally (Figure 1(a)). Any change in the device environment (in this case dielectric environment with dielectric constant, $\epsilon_c$) that alters either scattering contribution will be reflected as a shift in $T_t$ (Figure 1(a)). We use this approach to investigate the effect of HfO$_2$ capping on single-layer MoS$_2$.

We present measurements on two types of devices: one without HfO$_2$ and another capped with HfO$_2$. Figure 1(b) shows a schematic of a single-layer MoS$_2$ device fabricated on a 285 nm SiO$_2$ layer on top of a heavily doped silicon substrate, which serves as a global back gate ($V_{bg}$). Gate-voltage-dependent transconductance traces as a function of temperature are provided in Supplementary Figure S1. A pronounced shift in threshold voltage is observed as the temperature decreases, consistent with the suppression of thermionic emission. From the linear region of the transfer curves, we extract the field-effect mobility using

$$\mu = LWC_{ox}\left(\frac{dG}{dV_{bg}}\right) \quad (1)$$

where $L$ and $W$ are the channel length and width, and $C_{ox}$ is the back-gate capacitance. The extracted temperature-dependent mobility is shown in Figure 1(c). In single-layer MoS$_2$, the crossover temperature $T_t$ is found to be ~330 K. Fitting the high-temperature region to the generic power-law dependence $\mu_{ph} \sim T^{-\gamma}$ depends on the dominant phonon scattering mechanism, yields $\gamma = 2.2$, consistent with a previous report [20].

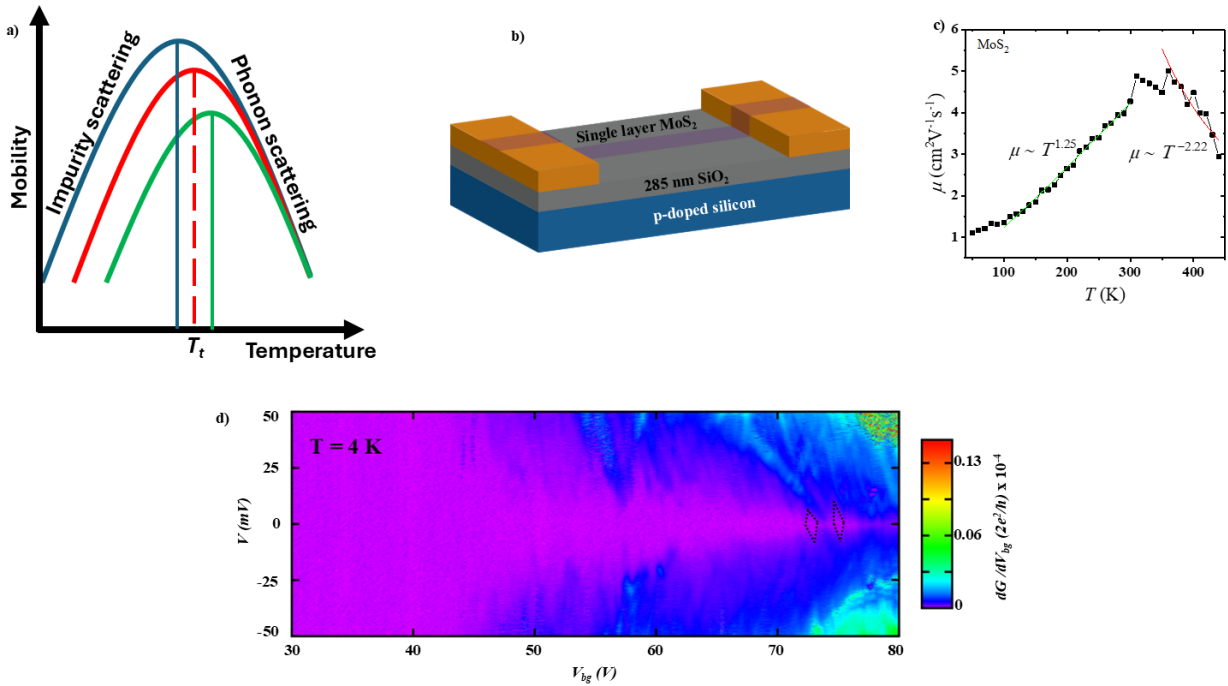

*Figure 1. (a) Schematic illustration of temperature-dependent mobility trends. At high temperatures, mobility is limited by phonon scattering, whereas at low temperatures it is limited by impurity scattering. The crossover region marks the temperature where the two contributions are comparable. The three traces depict how the crossover temperature shifts with varying impurity and phonon scattering strengths. (b) Schematic of the single-layer MoS$_2$ device fabricated on a SiO$_2$/Si substrate without a high-k dielectric (HfO$_2$) capping layer. (c) Temperature dependence of electron mobility in the single-layer MoS$_2$ device without HfO$_2$ capping. (d) 2D conductance map as a function of back-gate voltage and source–drain voltage at T = 4 K.*

At temperatures $T \leq 30$ K, conductance shows a much weaker variation with gate voltage, indicating that transport is dominated by hopping through localized states [21], driving the system into a strongly localized regime. Figure 1(d) presents a conductance colormap as a function of source-drain bias ($V$) and $V_{\text{bg}}$. In several regions, we observe diamond-shaped domains where conductance is completely suppressed inside the diamond, consistent with transport governed by Coulomb blockade [22]. The confinement responsible for the Coulomb blockade is attributed to potential fluctuations and disorder originating from substrate roughness [23], [24].

We next turn to single-layer MoS$_2$ devices capped with a 30 nm HfO$_2$ layer using atomic-layer-deposited (ALD). The rest of the fabrication process remains identical to the uncapped devices. The schematic of the capped device is shown in Figure 2(a), and gate-voltage-dependent transconductance traces as a function of temperature are provided in Supplementary Figure S2. Unlike the uncapped device, the HfO$_2$-capped MoS$_2$ exhibits a metal–insulator transition at $T_t \sim 175$ K. Additionally, the threshold voltage is shifted to more negative values compared to the uncapped case. Two possible mechanisms may account for this shift. First, ALD grown HfO$_2$ can form chemical bonds at intrinsic point defects in MoS$_2$, leading to charge transfer equivalent to electrostatic doping [18]. Second, ALD growth can introduce defect states at the MoS$_2$/HfO$_2$ interface, which may pin the Fermi level and weaken the gate efficiency [18]. A third possibility is that both mechanisms contribute simultaneously. We denote the screening dielectric constant of the capping as $\epsilon_c$.

From the linear part of the transconductance traces, we extract the field-effect mobility as a function of temperature (Figure 2(b)). Compared to the uncapped device, the capped MoS$_2$ shows a shift of the crossover temperature $T_t$ to lower values, indicating a reduced contribution from impurity scattering, consistent with earlier experimental observations [16]. One can assume a generic power-law dependency for mobilities [25]: $\mu_{\text{imp}} \sim \epsilon_c^{\beta_\epsilon} T^\beta$ and $\mu_{\text{ph}} \sim \epsilon_c^{\gamma_\epsilon} T^{-\gamma}$, with $\beta_\epsilon$ and $\gamma_\epsilon$ are exponents that control the power-law dependence of impurity- and phonon-related mobility branches on the screening parameter $\epsilon_c$, while $\beta$ and $\gamma$ describe the dependence on temperature $T$. The crossover temperature condition $\mu_{\text{imp}}(T_t) = \mu_{\text{ph}}(T_t)$ gives $T_t \sim \epsilon_c^{(\gamma_\epsilon - \beta_\epsilon)/(\beta + \gamma)}$. Mobility dependence on screening $\epsilon_c$ is complicated, nevertheless, $\mu_{\text{imp}}$ increases with $\epsilon_c$ typically with $\beta_\epsilon \approx 2$ [25], while $\mu_{\text{ph}}$ decreases with $\epsilon_c$ typically for MoS$_2$ with $\gamma_\epsilon \approx -0.6$ even to $-2$ (for higher $\epsilon$) [25]; that $\beta + \gamma$ is always positive implies that $T_t$ decreases with $\epsilon_c$, which we observe.

At $T < 30$ K, pronounced conductance oscillations emerge (Supplementary Figure S2). When plotted as a conductance colormap versus both $V$ and $V_{\text{bg}}$, these oscillations evolve into a checkerboard-like pattern. Such patterns are a hallmark of Fabry-Pérot (FP) interference, arising when electrons traverse the channel and are partially reflected between scattering centers, forming standing wave resonances [26]. The checkerboard energy scale is inversely proportional to the cavity length, with the energy spacing of FP resonances (for a single mode channel with the quadratic dispersion relation) given by

$$\Delta E(n) = \frac{\hbar^2 \pi^2}{2m^* L_c^2}(2n - 1), \qquad (2)$$

where $m^* = 0.4\, m_e$ [27] is the effective electron mass in MoS$_2$ and $L_c$ is the effective cavity length. From the observed energy range in Figure 2(c), we extract cavity lengths on the order of 30-100 nm. The coexistence of FP resonances with different periodicities further suggests that multiple cavities contribute to the phase-coherent transport. The observation of FP interference requires that the phase-coherence length exceeds the mean free path, thereby providing unambiguous evidence of quantum-coherent transport in HfO$_2$-capped MoS$_2$. Most of the fine conductance oscillations that originate due to FP resonances fade away at temperatures between 6 K to 12 K (Fig. 2(d)).

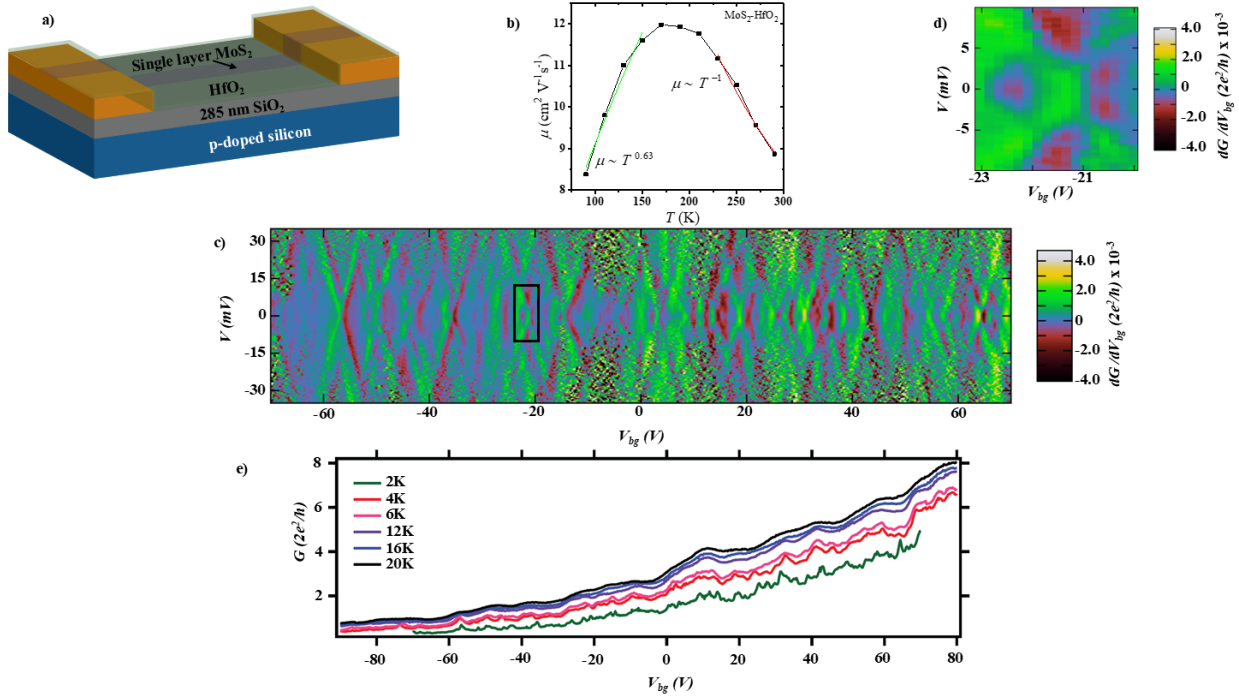

*Figure 2. (a) Schematic of the single-layer MoS2 device fabricated on a SiO2/Si substrate with high-k dielectric (HfO$_2$) capping layer. (b) Temperature dependence of electron mobility in the single-layer MoS$_2$ device with HfO$_2$ capping. (c) 2D conductance map as a function of back-gate voltage and source–drain voltage at T = 4 K.(d) Zoomed-in region (black square in panel c) displaying conductance checker-board pattern, a hallmark of Fabry-Pérot interference. (e) Transconductance traces at various temperatures, showing the disappearance of fine oscillations that evolve into the checkerboard pattern observed in panel (c).*

# Discussion

In our devices, the distinction between classical and quantum diffusive regimes is evident when comparing uncapped and HfO$_2$-capped single-layer MoS$_2$. In the uncapped device, the phase coherence length ($l_\varphi$) is shorter than or comparable to the mean free path ($l$): $l_\varphi < l$, such that scattering randomizes both momentum and phase. This places the device in the classical diffusive regime, where transport is governed by incoherent single-electron tunneling and no interference features are observed. By contrast, in the HfO$_2$-capped device we find $l < l_\varphi < L$, indicating that electrons maintain their quantum phase coherence across multiple scattering events. This condition enables the emergence of FP interference, where coherent electron waves reflect between scattering centers to form standing-wave patterns. The observation of such interference directly demonstrates that dielectric engineering with HfO$_2$ drives the system into the quantum diffusive regime, where transport is dictated by wave nature rather than classical particles.

To quantitatively establish the influence of the high-κ dielectric on screening effects and the suppression of impurity scattering, we develop an interferometer-based device model.

## Interferometer device model

The model of an interferometer device based on TMD monolayer is presented in Fig. 1. We start with the defect-free monolayer material represented by a rhombus flake of size $N_x \times N_y = 480 \times 120$ unit cells, schematically shown in Fig.3(a), top. The first dimension, $N_x$, defines the device (channel) length, here $L \simeq 160$ nm. To simplify the transport description, we freeze the movement in the lateral $y$ direction by a gaussian potential profile $U^{ch}(x,y) = U_0(1 - \exp[-y^2/2\sigma])$, with $U_0 = -120$ mV and $\sigma = 5$ nm. The potential profile is presented in Fig. 3(a), bottom, forming a conductive channel along the $x$ direction.

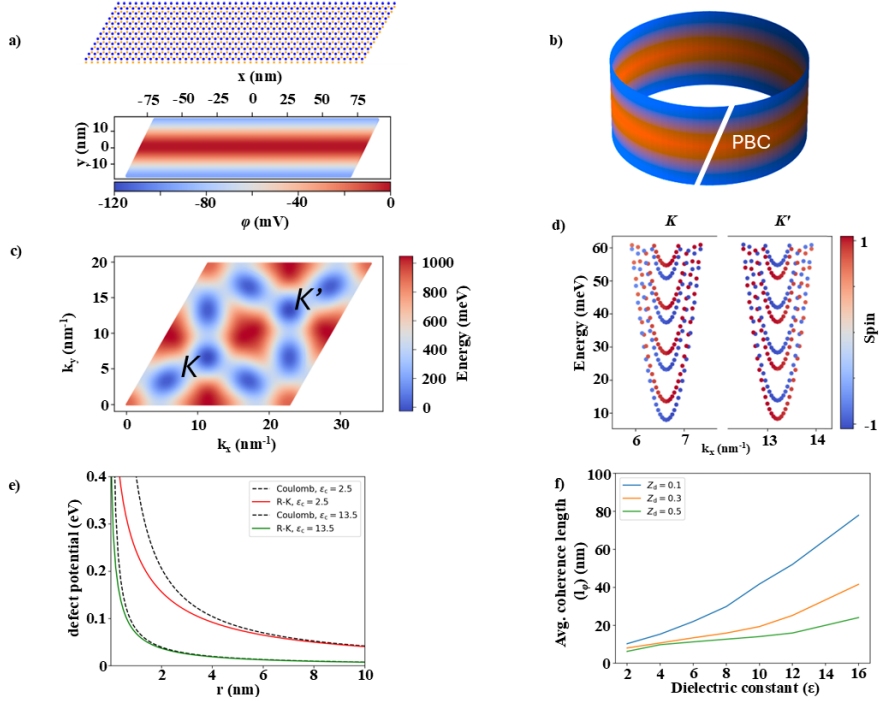

*Figure 3. Single-particle tight-binding theory for (a) finite flake of size $N_x \times N_y = 480 \times 120$ unit cells with applied gate-defined channel potential together with (b) periodic boundary conditions (PCB) were used to describe (c) band structure – presented CB minimum with marked K/K' valleys, and (d) calculate conductive modes which then enter the Landauer formula. (e) Defect Rytova-Keldysh potential describing screened charge within thin slab capped by (different) dielectric environment. Two capping screenings assumed ($\epsilon_c = 2.5$ and $13.5$) correspond to the two experimental setups. For comparison respective Coulomb potential also added. (f) Average coherence length as a function of defects screening realized by dielectric capping $\epsilon_c$.*

F-P interference in nanoscale devices can be modeled using the non-equilibrium Green's function (NEGF) approach [28] or the S-matrix formalism [26], [29], which, in the simplest two-scatterer configuration, yields explicit formulas for the transmission matrix [28], [30]. In this work, however, we adopt a different approach: we directly construct the conductive channel modes as eigenstates of the tight-binding (TB) Hamiltonian with periodic boundary conditions (PBC) applied along the channel direction. Within this framework, the eigenstates of a channel of length, $L$, take the form of Bloch waves, with quantized wave vectors given by $k_{ch} = 2n\pi/L$. This quantization is equivalent to the FP cavity quantization, $k_{FP} = n\pi/L_c$, where the cavity length $L_c = L/2$. After obtaining the conductive modes, we evaluate the channel conductance using the Landauer–Büttiker formalism [30], [31].

## Tight-binding theory

Here we present the TB theory for MoS$_2$ rhombus flake present in the device. The nanostructure Hamiltonian for a single-electron state $i$ satisfies the Schrödinger equation:

$$(H_{\text{bulk}} - |e|\phi)|\Psi_i\rangle = E_i|\Psi_i\rangle. \tag{3}$$

The TB model for MoS$_2$ monolayer is expressed in a basis of five d-orbitals localized on Mo atoms and six p-like orbitals describing S$_2$ dimers (11 orbitals in total) [32]. We impose Born-von Karman periodic boundary conditions by wrapping the computational box on a torus, as illustrated in Fig. 3(b). This procedure yields a discrete set of allowed $k$-vectors, over which we diagonalize the bulk Hamiltonian $H_{\text{bulk}}$. In the subsequent step, the channel wavefunction $|\Psi_i\rangle$ is expanded in terms of the eigenstates of $H_{\text{bulk}}$ at each permitted $k$-value, and Eq. (3) is solved with the inclusion of the potential term $\phi$. Additional computational details are provided in Refs. [33], [34]. The conductive channel profile represented by $\phi(x,y) = U^{\text{ch}}(y) + \phi_{\epsilon_c}^{\text{def}}(x,y)$ field is depicted in Fig. 4(a-c). The defect potential part $\phi_{\epsilon_c}^{\text{def}}(x,y)$ is composed of $V_{\text{R-K}}(\epsilon_c r)$, modelled by the Rytova-Keldysh (R-K) potential, defined below in Eq. (4), localized at random positions within the channel – see Figs. 4(b,c). Defects set-up and their screening $\epsilon_c$ is discussed next.

## Defects modelling

We introduced defects to the system by modifying the on-site potential $\phi$ in the TB Hamiltonian, Eq. (3). Following DFT calculations (random-phase approximation (RPA)-screened impurity potential) [35], here each defect is modelled using the well-established Rytova-Keldysh (R-K) potential [36]. The R-K potential accounts for the nonlocal screening of a charge carrier in a thin dielectric film, explicitly incorporating the dielectric mismatch with the surrounding environment. In our case, the effective screening is described using the average dielectric constant of the top ($\epsilon_t$) and bottom ($\epsilon_b$) "capping" layers, given by $\epsilon_c = (\epsilon_t + \epsilon_b)/2$ [36], [37], [38]:

$$V_{\text{R-K}}(\epsilon_c r) = -Z_d U_0 \frac{\pi}{2}\left(H_0\left(\frac{\epsilon_c r}{r_0}\right) - Y_0\left(\frac{\epsilon_c r}{r_0}\right)\right), \quad (4)$$

with $H_0$ and $Y_0$ being Struve, and Bessel function of the second kind, respectively. We adopted from [35] the following parameters for the R-K potential: length-scale $r_0 = 5$ nm, and strength $U_0 = 1$ eV (defect located $d = 2$ Å above 2D plate) to correctly model defects in MoS$_2$ material. In the capped (Fig. 2(a)) device configuration where MoS$_2$ sits on SiO$_2$ at the bottom and capped with HfO$_2$ on the top), the $\epsilon_c = (\epsilon_t + \epsilon_b)/2 = (23 + 4)/2 = 13.5$. In the uncapped device configuration (Fig. 1(b)) where the MoS$_2$ sits on SiO$_2$ on bottom and vacuum on top, the $\epsilon_c = (1 + 4)/2 = 2.5$. The resulting defect potential $V_{\text{R-K}}(r)$ for these two screening regimes is presented in Fig. 3(e). Let us assume a defect in the form of charged acceptor adatom. $Z_d$ is an amount of charge transfer from the adatom to monolayer MoS$_2$. Typically, the transfer of charge on defect site is $Z_d < 1$, depending on defect type. Here, for defects localized above the Mo-site we estimate $Z_d = 0.3$ (estimated using RPA with Kohn-Sham wave functions from ab initio calculations [35], [39]). For comparison, in Fig. 3(e), we also present Coulomb potential profile $V_C(r) = -(Z_d U_0)/(\epsilon_c r)$. One can observe that screening effect in 2D structures can dramatically change impurities profile and thus their effective dimension (interaction length scale). In Fig 3(e) the $V_{\text{R-K}}(\epsilon_c r)$ with $\epsilon_c = 13.5$ (green curve) is far more localized than weakly screened defect potential $\epsilon_c = 2.5$ (red curve).

We assume an acceptor defect density of approximately $10^{12}$ cm$^{-2}$ [40], [41] and distribute them randomly on the monolayer surface, as illustrated in Fig. 4(b,c). For the simulated channel area, this corresponds to 30 defect centers in total, consistent with the realistic defect density of our device. It is important to note that, because the defects are randomly distributed in our model, they do not form ordered structures such as waveguides observed in photonic crystals [42], [43].

## Coherence length estimation

After introducing defects into the system, a key step in describing transport is to estimate the electron coherence length, $l_\varphi$. In the absence of defects, the coherence length is limited by the system size, i.e. $l_\varphi \simeq L_c$. When defects are present, $l_\varphi$ is reduced as conductive modes begin to localize around defect sites. We estimate the coherence length as the spatial width of the localized state using the uncertainty relation for wave packets, $\sigma_x \sigma_k = 1/2$, which yields $l_\varphi = 1/(2\sigma_k)$ where $\sigma_k$ is the standard deviation of the wavevector component $k_x$ along the channel direction for a given eigenstate. For the chosen defect configuration in the channel (Fig. 4(b,c)), we compute the eigenstates $|\Psi_i\rangle, i = 1, \ldots, 200$ (the 200 lowest states) of the TB Hamiltonian (Eq. (3)) and evaluate the average coherence length $\langle l_\varphi \rangle$ over this set. The procedure is repeated for different dielectric screening environments, and the results are shown in Fig. 3(f).

We find that increasing the dielectric screening $\epsilon_c$ of the defect potential landscape (Fig. 4(b,c)), for various assumed defect charges $Z_d$, leads to a systematic increase in the average coherence length. For smaller defect charges (blue and orange curves), once $\epsilon_c$ exceeds $\sim 10$, the coherence length $l_\varphi$ becomes comparable to the cavity size ($L_c \simeq 80$ nm in the TB simulations), thereby enabling coherent transport.

## Landauer formula for FP interference

After calculating the conductive modes within the TB model under periodic boundary conditions and estimating the coherence length for each mode, we proceed to evaluate the conductance of the defective channel for different screening strengths. The standard framework for transport calculations is provided by the Landauer-Büttiker formalism [30], [31], in which the linear-response conductance of a mesoscopic sample is given by [28]:

$$G(E) = \frac{e^2}{h} \sum_n \Theta(E - E_n) T_n(E), \qquad (5)$$

where $T_n$ denotes the transmission probabilities through the subbands of energy $E_n$, and $\Theta$ is the Heaviside step function, i.e., $\Theta(x) = 1$, for $x \geq 0$, and $\Theta(x) = 0$ otherwise.

In our case, the eigenstates are discrete due to the finite channel size, and thus the density of states for each subband can be approximated as $\Theta(E - E_n) = \sum_k \delta(E - E_{nk})$ where $E_{nk}$ describes the dispersion of subband $n$, and the summation runs over discrete conductive modes $k$. To make the calculations more realistic, we account for level broadening [28], [29] by replacing the delta function $\delta(E - E_{nk})$ with a Gaussian profile $g \exp[-(E - E_{nk})^2/(2\sigma_E^2)]$, where the amplitude $g$ and dispersion $\sigma_E$ are tuned to reproduce both the amplitude and broadening of the experimental FP oscillations.

A second important assumption is that, for a defective channel, the transmission should be reduced below unity. Here we estimate the transmission probability as $T_{nk}$ as a ratio $l_\varphi(n,k)/L$ [44], i.e., the ratio of the coherence length of mode $E_{nk}$ to the channel length $L$. For $l_\varphi \simeq L$, the transmission approaches the ballistic limit. To relate the gate voltage $V_{bg}$ to the chemical potential $E$, we use a gate lever arm of $\alpha = 6 \times 10^{-4}$, such that $E = \alpha V_{bg}$. Finally, to incorporate the conductance dependence on the source-drain bias $V$ we follow a method described in [29].

The TB simulation results for conductance in the defective channel are shown in Fig. 4. Subfigures 4(a-c) present the channel potential landscapes corresponding to three different screening setups: (a) no defects, (b) weakly screened defects (as in experimental setup Fig. 1(b)), and (c) strongly screened defects (as in experimental setup Fig. 2(a)). The corresponding eigenstates of the TB Hamiltonian (Eq. (3)) are displayed in Figs. 4(d-f).

In the defect-free case (Fig. 4(d)), the eigenstates form characteristic subbands that are spin-split (red and blue dots denote spin orientations) due to the strong spin-orbit interaction inherent to TMDC materials. The subbands arise from lateral quantization within the modeled quantum channel, and the discretization of eigenstates in $k$-space ($k_x$ is a good quantum number) reflects the finite channel length $L$. We can number each eigenstate by calculating $\langle k_x \rangle$.

When defects are introduced (Fig. 4(e)), the subband structure is destroyed: $\langle k_x \rangle$ changes dramatically, and $k_x$ ceases to be a good quantum number and stops characterizing subsequent modes in the channel. In this case, $\langle k_x \rangle$ is only meaningful as an expectation value for each mode. Remarkably, introducing a strong dielectric capping layer (Fig. 4(f)) partially restores the subband structure, thereby enabling coherent transport.

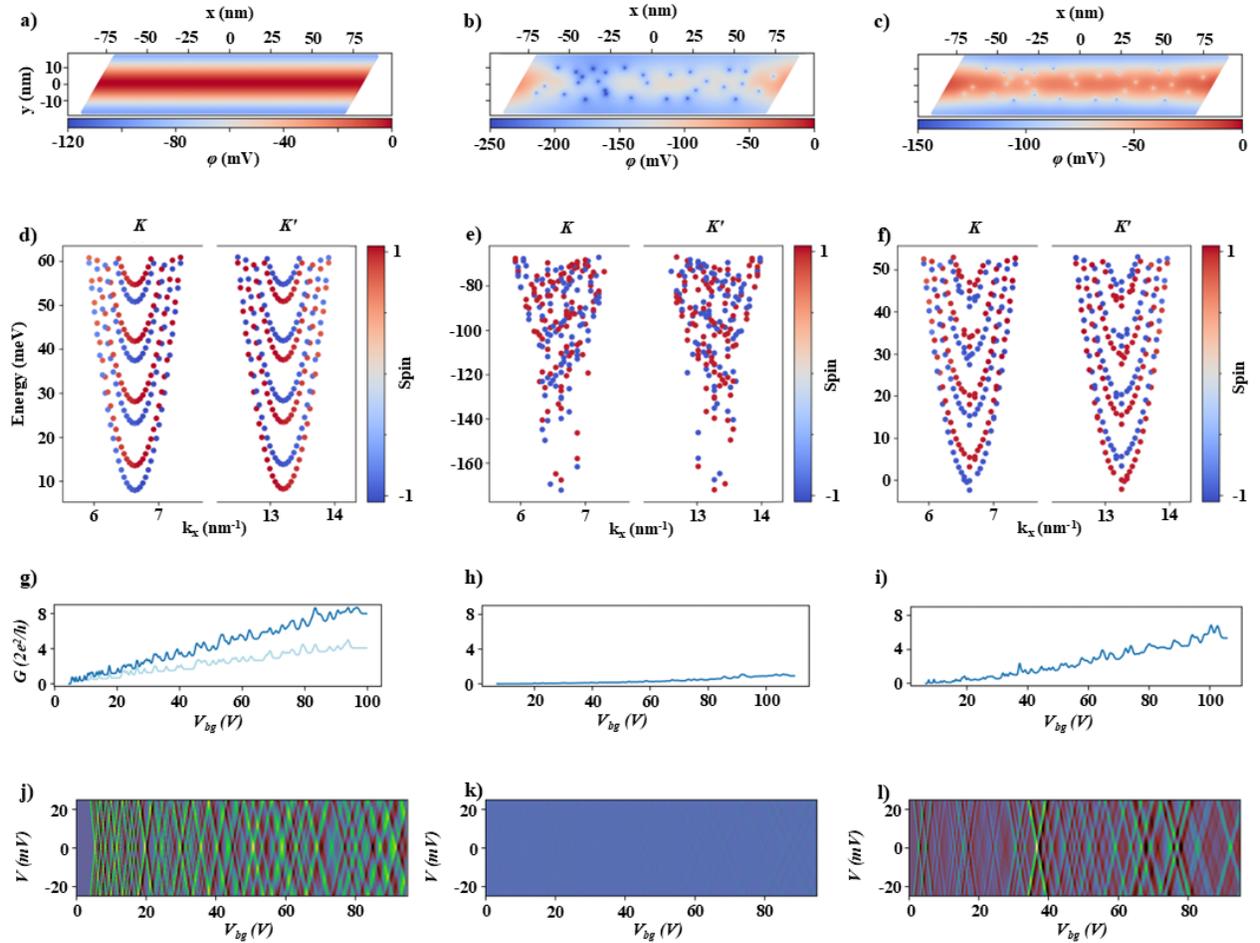

*Figure 4. The three transport regimes simulated in the F-P interferometer device: (a,d,g,j) No defects (ballistic transport). (b,e,h,k) Transport suppressed by defects (located in random positions and each modelled as $Z_d = 0.3$, $\epsilon_c = 2.5$). (c,f,i,l) Transport restored by screening of defects ($Z_d = 0.3$, $\epsilon_c = 13.5$). First row (a,b,c) presents the conductive channel potential lanscape for the three scenarios: (a) defect-free channel, (b) defective channel, and (c) with screened defects. Second row (d,e,f) present respective spectra. Third row (g,h,e) shows calculated respective conductances; and fourth one (j,k,l) conductance maps for different offsets $V_{bg}$ and source-drain biases $V$ with FP patterns present in (j) defect-free and (l) screened-defect regimes.*

Having obtained the subsequent modes in our Fabry-Pérot devices, we use the Landauer-Büttiker formula (Eq. (5)) to estimate the channel conductance. The results for a defect-free channel are shown in Figs. 4(g,j), where ballistic transport is observed. Small variations in the parameters $g$ and $\sigma_E$ can lead to a more staircase-like, quantized conductance, as illustrated by the light-blue curve in Fig. 4(g). By varying the gate offset $V_{\text{bg}}$ and source–drain bias $V$ in the simulations, we reproduce the characteristic FP interference patterns, also visible in Fig. 4(j).

For the defective channel (Figs. 4(h,k)), the transport is strongly suppressed, and the FP oscillations are almost completely quenched, as seen in Fig. 4(k). However, when the defects are screened by a strong dielectric environment (Figs. 4(i,l)), coherent transport is partially restored, leading to the reappearance of the FP patterns, as seen in Fig. 4(l), even in the presence of defects.

In the simulations, we set the cavity length to $L_c = L/2 = 80$ nm, which gives energy spacing, from Eq. (2), of $\Delta E(n) \simeq 0.037(2n-1)$ meV. For $n = 1\ldots 9$ (states located between the bottoms of the first and second channel modes), gives $\Delta E = 0.14\text{-}2.50$ meV and corresponds to $\Delta E_c = 0.54$ meV ($\alpha \times 0.9$ eV, with the gate lever arm $\alpha = 6 \times 10^{-4}$) extracted from the experimental data; see Fig. 2(d). In practice, however, since up to four channel modes contribute to the conductance in the simulations (Figs. 4(d-f)), the energy spacing between successive FP resonances becomes more complex, as shown in Figs. 4(j-l).

# Conclusion

In summary, we have demonstrated that the dielectric environment plays a decisive role in determining both classical and quantum transport in single-layer MoS₂ devices. By comparing uncapped devices with HfO₂-capped devices, we show that the presence of a high-κ dielectric suppresses impurity scattering and shifts the transport from a classical diffusive regime toward a quantum-coherent regime, as evidenced by the emergence of Fabry–Pérot interference patterns at low temperatures. Temperature-dependent mobility measurements reveal a reduction in impurity-limited scattering in HfO₂-capped devices, consistent with enhanced dielectric screening. Tight-binding interferometer simulations further confirm that the dielectric layer modifies the conductive modes and facilitates phase-coherent transport. These results highlight that careful engineering of the dielectric environment can be used not only to enhance device performance but also to control the quantum behavior of charge carriers in two-dimensional semiconductors. Our findings establish high-κ dielectrics as a versatile tool for realizing coherent quantum transport and provide a pathway for the design of future nano electronic and quantum devices based on transition metal dichalcogenides.

# Acknowledgement

This research is supported by MonArk NSF Quantum Foundry supported by National Science Foundation Q-AMASE-i program under NSF Award No. DMR-1906383 and AFRL under agreement number FA8750-24-1-1019. JP acknowledge support from National Science Centre, Poland, under grant no. 2021/43/D/ST3/01989.

# Supplementary Information for "Classical to Quantum Diffusive Transport in Atomically Thin Semiconductors Capped with High-k Dielectric"

*Jarosław Pawłowski [1], Dickson Thian[2], Repaka Maheswar[2], Chai Jian Wei[2], Pawan Kumar[2], Sudhiranjan Tripathy[2], Hugh Churchill[3,4,5], Dharmraj Kotekar Patil[3,4,5]*

[1] Institute of Theoretical Physics, Wrocław University of Science and Technology, Wrocław, Poland

[2] Institute of Materials Research and Engineering, Agency for Science Technology and Research, (A*STAR) Singapore 138634, Republic of Singapore 138634

[3] Institute for Nano Science and Engineering, University of Arkansas, Fayetteville, AR 72701 USA

[4] Department of Physics, University of Arkansas, Fayetteville, AR 72701 USA

[5] MonArk NSF Quantum Foundry, University of Arkansas, Fayetteville, AR 72701 USA

S1: Transconductance in uncapped $MoS_2$ devices as a function of temperature.

S2: Transconductance in $HfO_2$ capped $MoS_2$ devices as a function of temperature.

S3: Arrhenius plot and activation energy in uncapped and capped $MoS_2$ device.

S4: F-P interference in second device.

S5: F-P interference in third device.

S6: Transconductance traces as function of temperature.

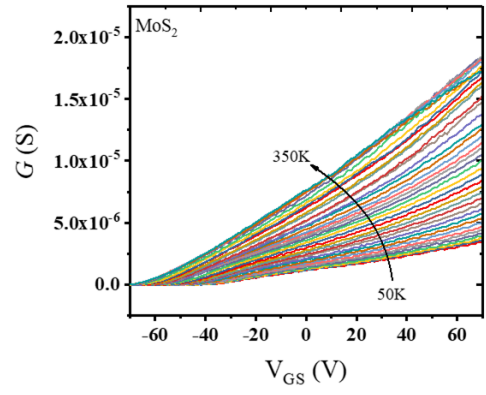

*Figure S1: Transconductance traces at various temperatures in uncapped MoS$_2$ device, used for extracting mobility in main text Fig. 1(c).*

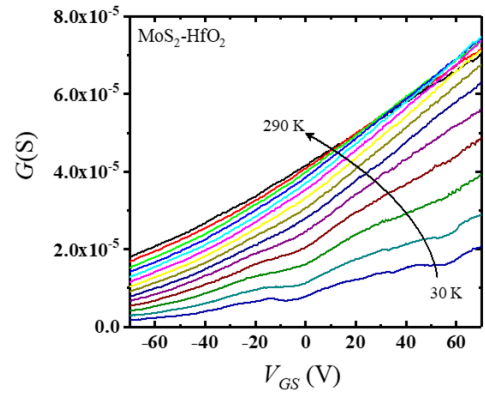

*Figure S2: Transconductance traces at various temperatures in HfO$_2$ capped MoS$_2$ device, used for extracting mobility in main text Fig. 2(b).*

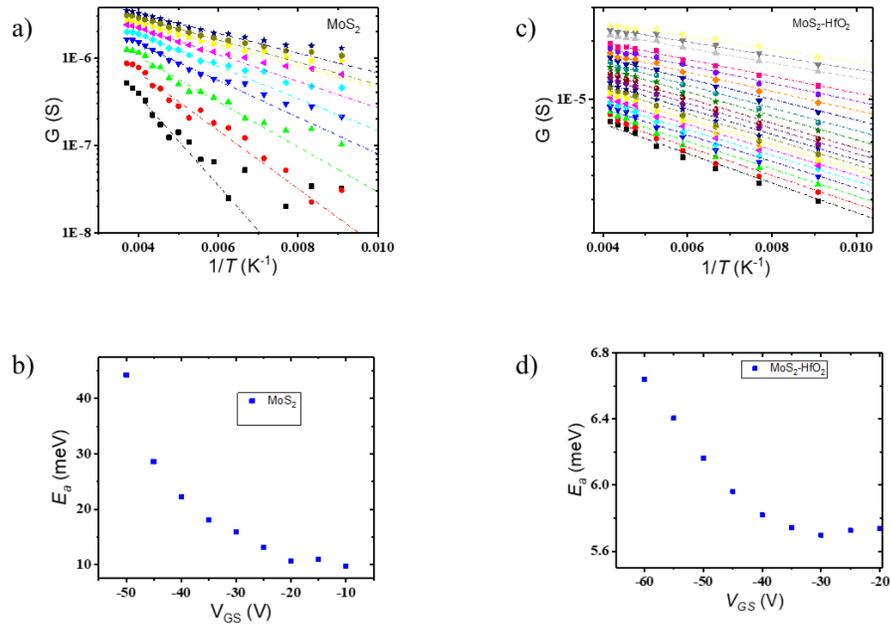

*Figure S3: Arrhenius plot in uncapped (a) and capped (c) and activation energy in uncapped (b) and capped (d) MoS$_2$ device.*

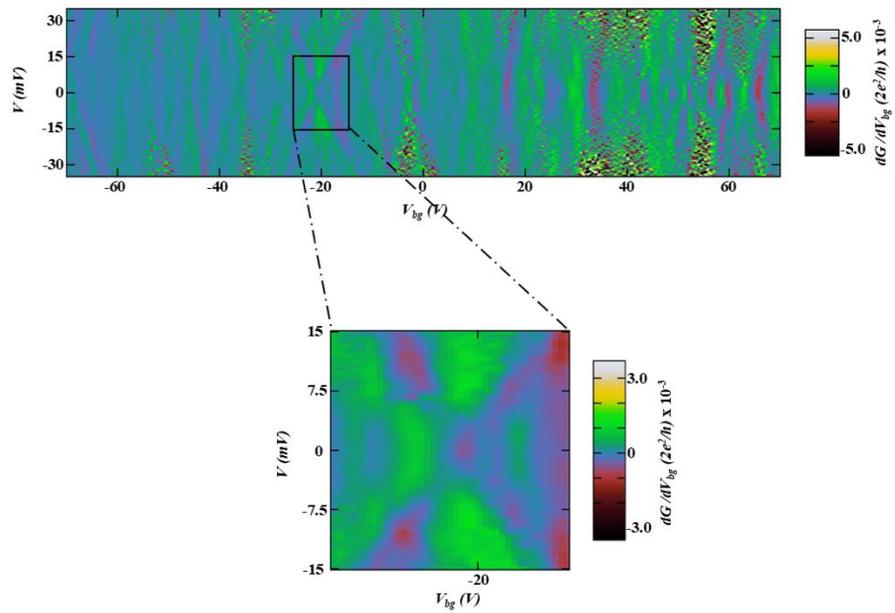

*Figure S4: 2D conductance map as a function of backgate and source-drain voltage exhibiting checker-board pattern in second device capped with HfO$_2$ on single layer MoS$_2$ device.*

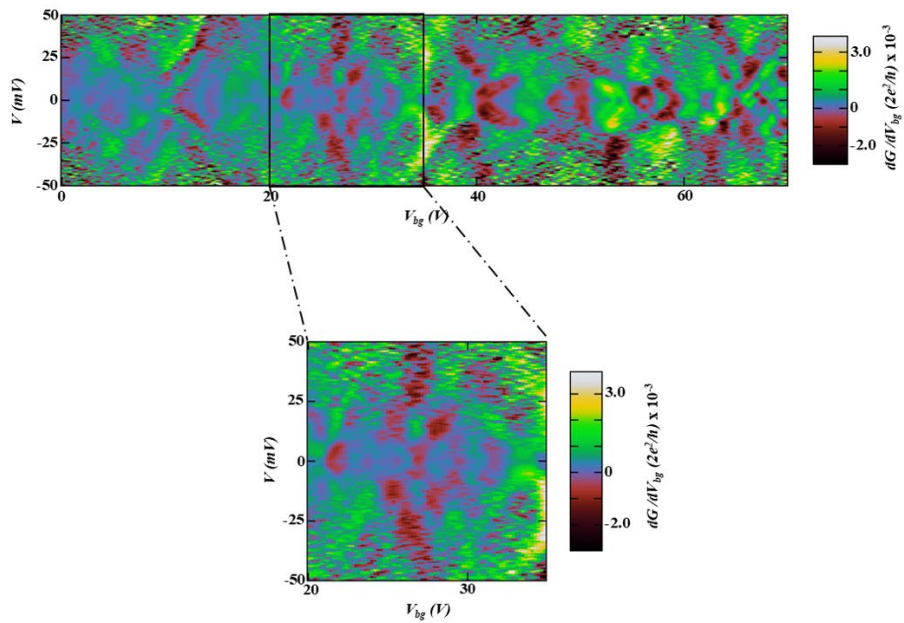

*Figure S5: 2D conductance map as a function of backgate and source-drain voltage exhibiting checker-board pattern in third device capped with $HfO_2$ on single layer $MoS_2$ device.*

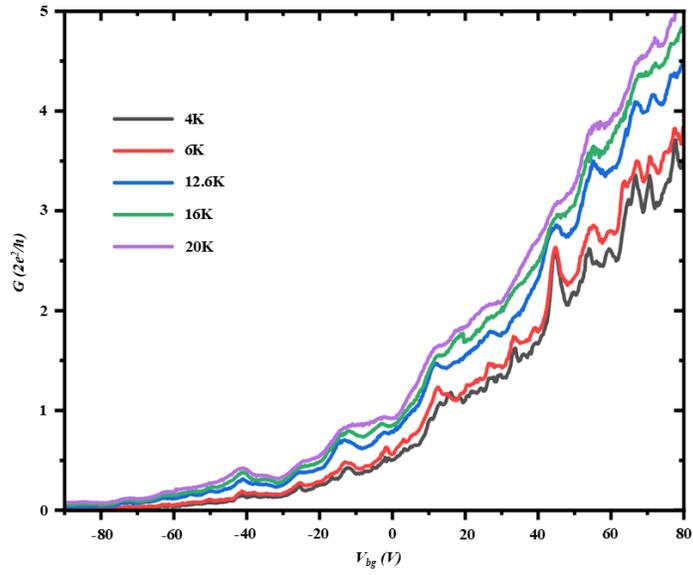

*Figure S6: Transconductance traces as a function of various temperature in $HfO_2$ capped $MoS_2$ device (second device, 2D conductance map shown in Fig. S4) where fine conductance oscillations which evolve into checkerboard pattern disappear between T = 6K to 12K.*